\documentclass[a4paper]{jpconf}

\usepackage{graphicx}
\usepackage{url}
\usepackage{lineno}

\begin{document}

\title{Neutron and proton tests of different technologies for the upgrade of the cold readout electronics of the ATLAS Hadronic End-cap Calorimeter}

\author{M Nagel}

\address{Max-Planck-Institut f\"{u}r Physik (Werner-Heisenberg-Institut), F\"{o}hringer Ring 6,\\80805 Munich, Germany, \url{www.mpp.mpg.de}}

\ead{nagel@mpp.mpg.de}

\author{on behalf of the HECPAS Collaboration\\
(IEP Ko\v{s}ice, Univ. of Montr\'{e}al, MPI Munich, IEAP Prague, NPI \v{R}e\v{z})}


\begin{abstract}
The expected increase of total integrated luminosity by a factor ten at the HL-LHC compared to the design goals for LHC essentially eliminates the safety factor for radiation hardness realized at the current cold amplifiers of the ATLAS Hadronic End-cap Calorimeter (HEC). New more radiation hard technologies have been studied: SiGe bipolar, Si CMOS FET and GaAs FET transistors have been irradiated with neutrons up to an integrated fluence of $2.2 \cdot 10^{16} \, \rm{n}/ \rm{cm^2}$ and with 200 MeV protons up to an integrated fluence of $2.6 \cdot 10^{14} \, \rm{p}/ \rm{cm^2}$. Comparisons of transistor parameters such as the gain for both types of irradiations are presented.
\end{abstract}

\section{The ATLAS Hadronic End-cap Calorimeter}

\begin{figure}\label{LAr-cryostat}
  \begin{center}
    \includegraphics[scale=0.7]{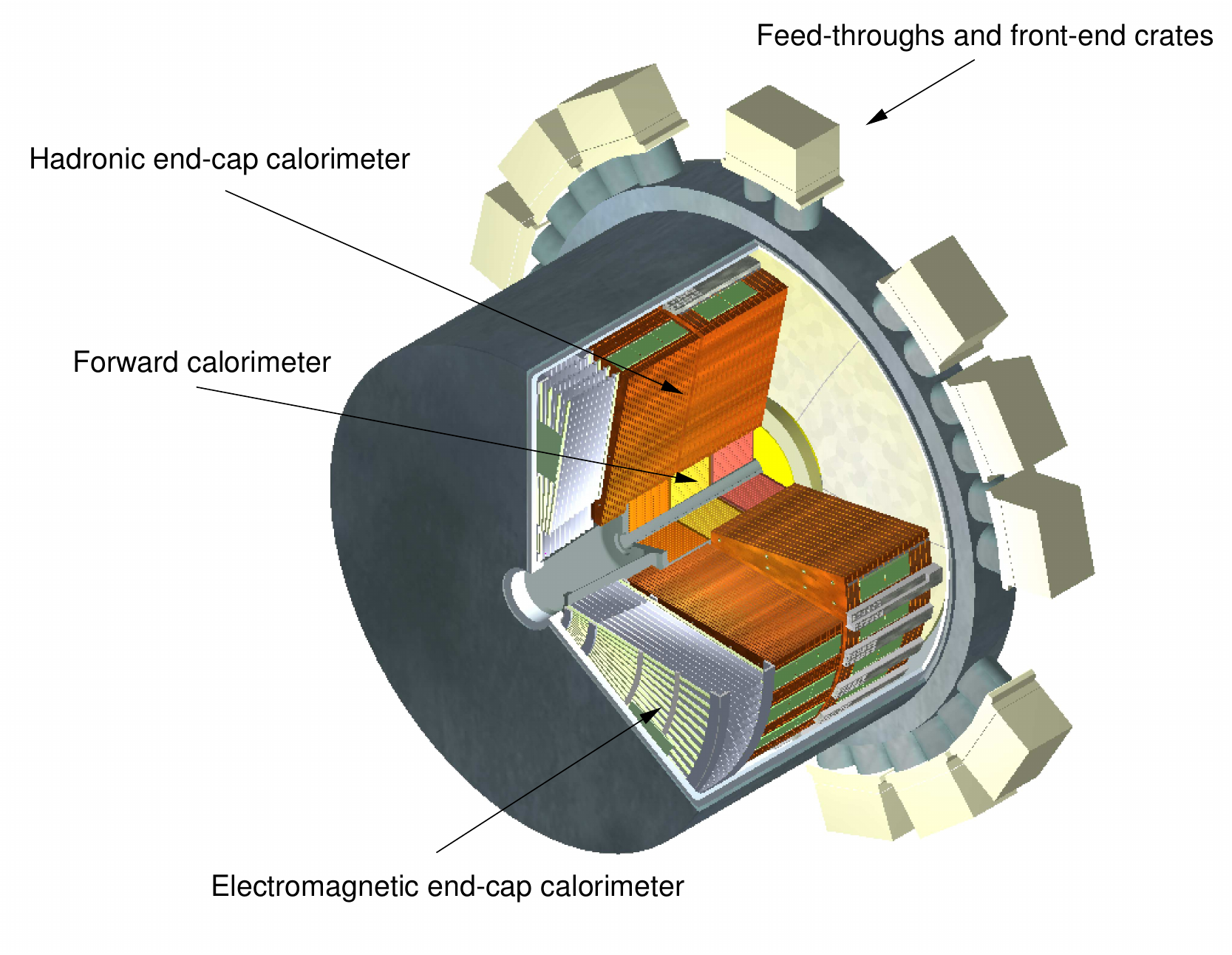}
    \caption{A liquid Argon end-cap cryostat, containing the hadronic and electromagnetic end-cap calorimeters, and the forward calorimeter~\cite{JINST_3_S08003}.}
  \end{center}
\end{figure}

The hadronic end-cap calorimeter (HEC) of the ATLAS experiment~\cite{ATLAS,JINST_3_S08003}
at the CERN Large Hadron Collider (LHC) is a copper-liquid argon
sampling calorimeter in a flat plate design~\cite{LArTDR,MPP-2007-237}. The
calorimeter provides coverage for hadronic showers in the
pseudorapidity range $1.5 < |\eta| < 3.2$. The HEC shares each of the
two liquid argon end-cap cryostats with the electromagnetic end-cap
(EMEC) and forward (FCAL) calorimeters, and consists of two wheels per
end-cap, as illustrated in Figure \ref{LAr-cryostat}.

A HEC wheel is made of 32 modules, each with 40 liquid argon
gaps, which are instrumented with active read-out pads. The signals
from the read-out pads are sent through short coaxial cables to
preamplifier and summing boards (PSB) mounted on the perimeter of the
wheels inside the cryostat. The PSB boards carry highly-integrated
preamplifier and summing amplifier chips in Gallium-Arsenide (GaAs)
technology. The signals from a set of preamplifiers are summed to one
output signal, which is transmitted to the cryostat
feed-through~\cite{MPP-2005-193}. Figure 2 shows a PSB board
mounted and connected on the perimeter of a HEC wheel.

\begin{figure}\label{PSB}
  \begin{center}
    \includegraphics[scale=0.38]{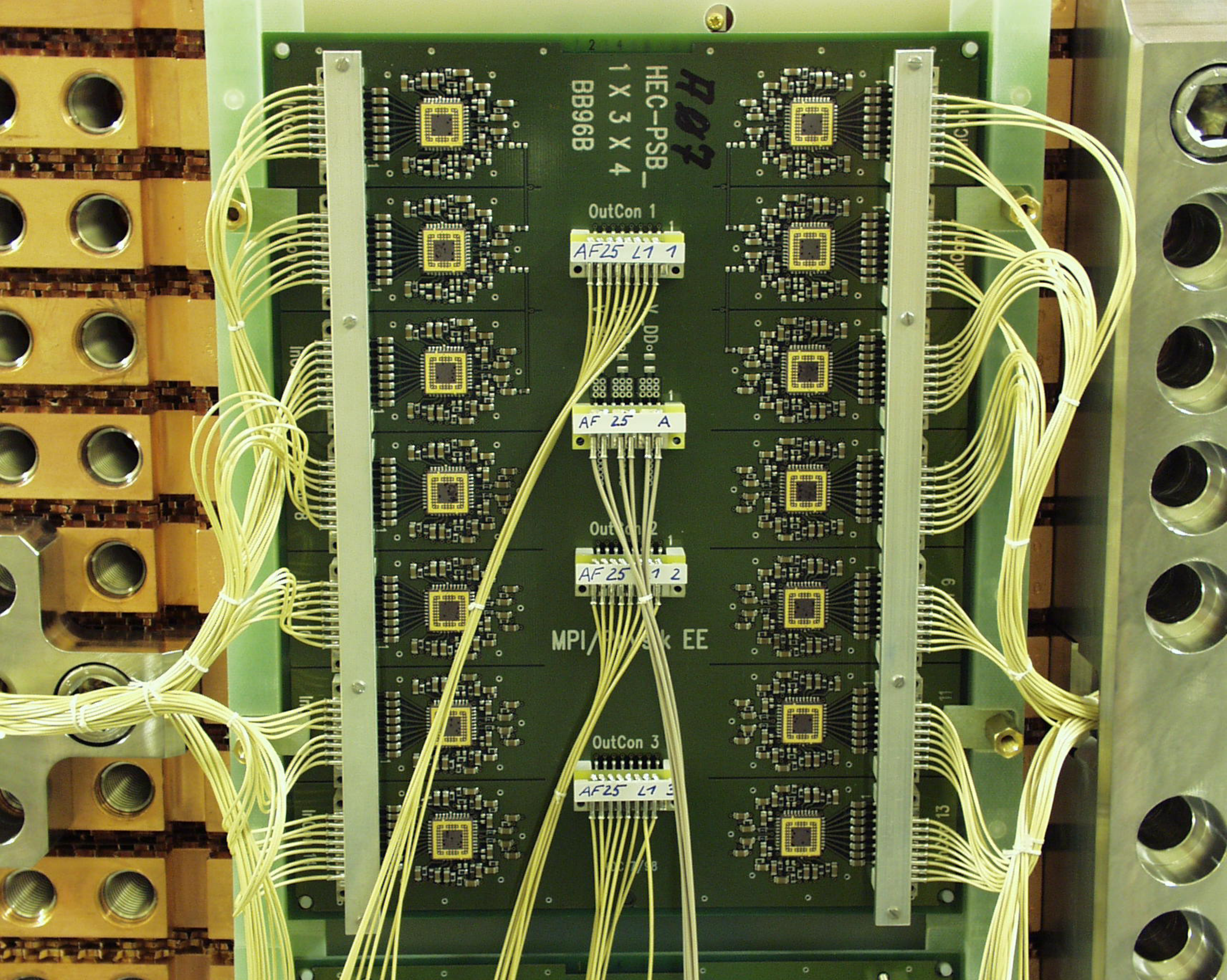}
    \caption{A detailed view of a PSB board mounted on the perimeter of a HEC wheel~\cite{MPP-2005-193}.}
  \end{center}
\end{figure}

\section{Requirements of the HEC cold electronics for the HL-LHC
  upgrade}

The GaAs technology currently employed in the HEC cold electronics has
been selected for its excellent high frequency performance, stable
operation at cryogenic temperatures, and radiation hardness. The
radiation hardness specifications were defined for ten years of
operation at the LHC design luminosity of $10^{34} \, \rm{cm^{-2} \, s^{-1}}$,
including a safety factor of ten. For the high-luminosity upgrade
of the LHC (HL-LHC), the luminosity is foreseen to increase by a
factor of 5--10, effectively eliminating the safety factor. The ATLAS collaboration
therefore decided to re-examine the radiation hardness of the current
HEC cold electronics and of potential alternative
technologies~\cite{Atlas-Upgrade,Schacht}. Detailed studies of the expected radiation levels after
ten years of running under HL-LHC conditions yielded the
following requirements (including a safety factor of 10) for the HEC cold electronics~\cite{ATL-GEN-2005-001}:

\begin{itemize}
\item 1 MeV equivalent neutron fluence (NIEL in Si) of $1.7 \cdot 10^{15} \, \mathrm{n} / \rm{cm}^{2}$
\item Total hadron fluence (SEE) of $1.8 \cdot 10^{14} \, \mathrm{h} / \rm{cm^{2}}$ with $E_h \, >$ 20 MeV
\item Total ionization dose (TID) of 22 kGy
\end{itemize}

\section{Tests}

The neutron irradiation tests were performed at the Fast Neutron Facility at the Nuclear Physics Institute of the ASCR in \v{R}e\v{z} near Prague, Czech Republic, up to an integrated fluence of $2.2 \cdot 10^{16} \, \mathrm{n} /
\rm{cm}^{2}$ (in terms of the 1 MeV equivalent NIEL in Si described in the next section). A $37\, \rm{MeV}$ proton beam incident on a $\rm{D_2O}$
target created a beam of neutrons with a continuous energy spectrum up to a maximum neutron energy of $34\, \rm{MeV}$ and a flux
density up to $10^{11} \, \mathrm{n} / \rm{cm^{2}} / \rm{s}$ with a $1/r^2$ decrease in flux density. The
proton irradiation test were performed at the Proton Irradiation
Facility at the Paul Scherrer Institute in Villigen, Switzerland, with a monoenergetic beam of $200\,
\rm{MeV}$ protons up to an integrated fluence of $2.6 \cdot
10^{14} \, \mathrm{p} / \rm{cm}^{2}$. The transistors were bonded in 
ceramic casings and mounted on boards, which were then placed in a standard aluminum frame with 17 mm distance between slots and carefully aligned in the 
particle beams. The three different transistor technologies being tested 
were Si CMOS FET in SGB25V 250nm technology from IHP, SiGe Bipolar HBT (IHP SGB25V 250nm and IBM 8WLBiCMOS 130nm), and the GaAs FET currently used in ATLAS, either the Triquint CFH800 250nm transistors themselves or integrated into the HEC BB96 Preamplifiers and Systems. The performance of 
the transistors was monitored \textsl{in situ} during irradiation with a
vector network analyzer recording a full set of $S$-parameters~\cite{SParameter}, which were converted to standard transistor parameters using suitable small signal circuit models.

\section{Flux determination}

A combination of radiation films and radmon diodes, placed at various slot positions along the beam, were used to determine the beam profile and, together with the beam current measurements, the particle flux. The films were exposed to a given beam current for various amounts of time and subsequently scanned.
An iterative calibration procedure was used to obtain the dose as a function of net optical density~\cite{RadfilmCalib}, by combining the known relative dose measurements with the absolute normalization obtained from MC simulations and from the radmon diodes.

For the neutron test with its continuous neutron energy distribution, the fluence was scaled to the 1-MeV equivalent neutron fluence in Silicon, employing the non-ionizing energy loss (NIEL) scaling hypothesis~\cite{NIEL}, independent of the actual technology under irradiation. The divergent nature of the neutron beam emerging from the heavy water target yielded a total fluence per slot given by $\Phi = 3.22 \cdot 10^{16} \cdot ({\rm slot\#} + \frac{3}{17})^{-2.11} \; {\rm n}/cm^2$. At PSI in contrast, the almost collimated proton beam together with an offset in the alignment of the boards yielded a basically constant fluence for all slots of $\Phi = (2.6 \pm 0.4) \cdot 10^{14} \; {\rm p}/cm^2$, where the main uncertainty stems from the unknown exact positions of the test structures inside the chip casing.


\section{Results}

\begin{figure}\label{Result_figure}
\begin{center}
\includegraphics[scale=0.6]{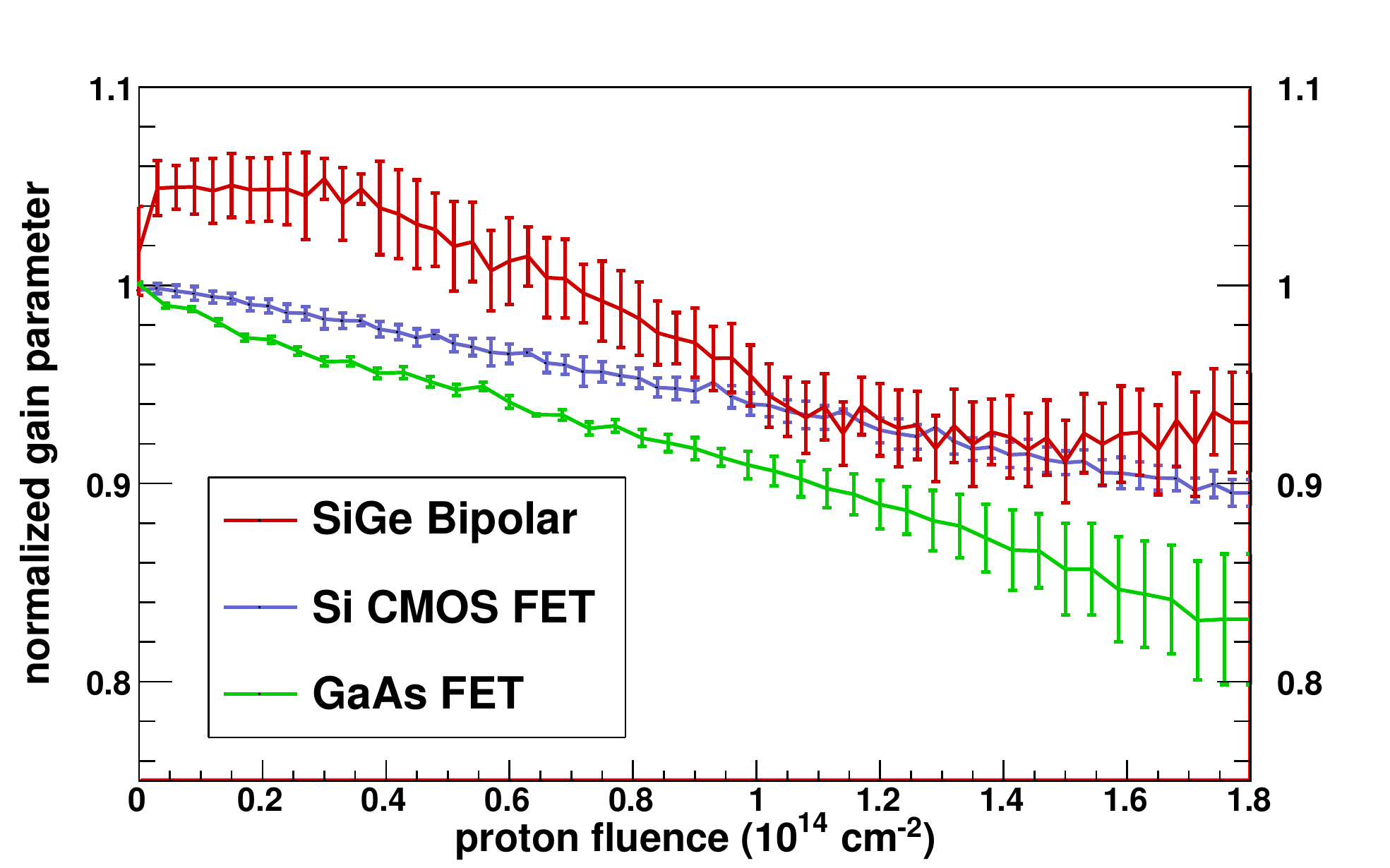}\\
\includegraphics[scale=0.6]{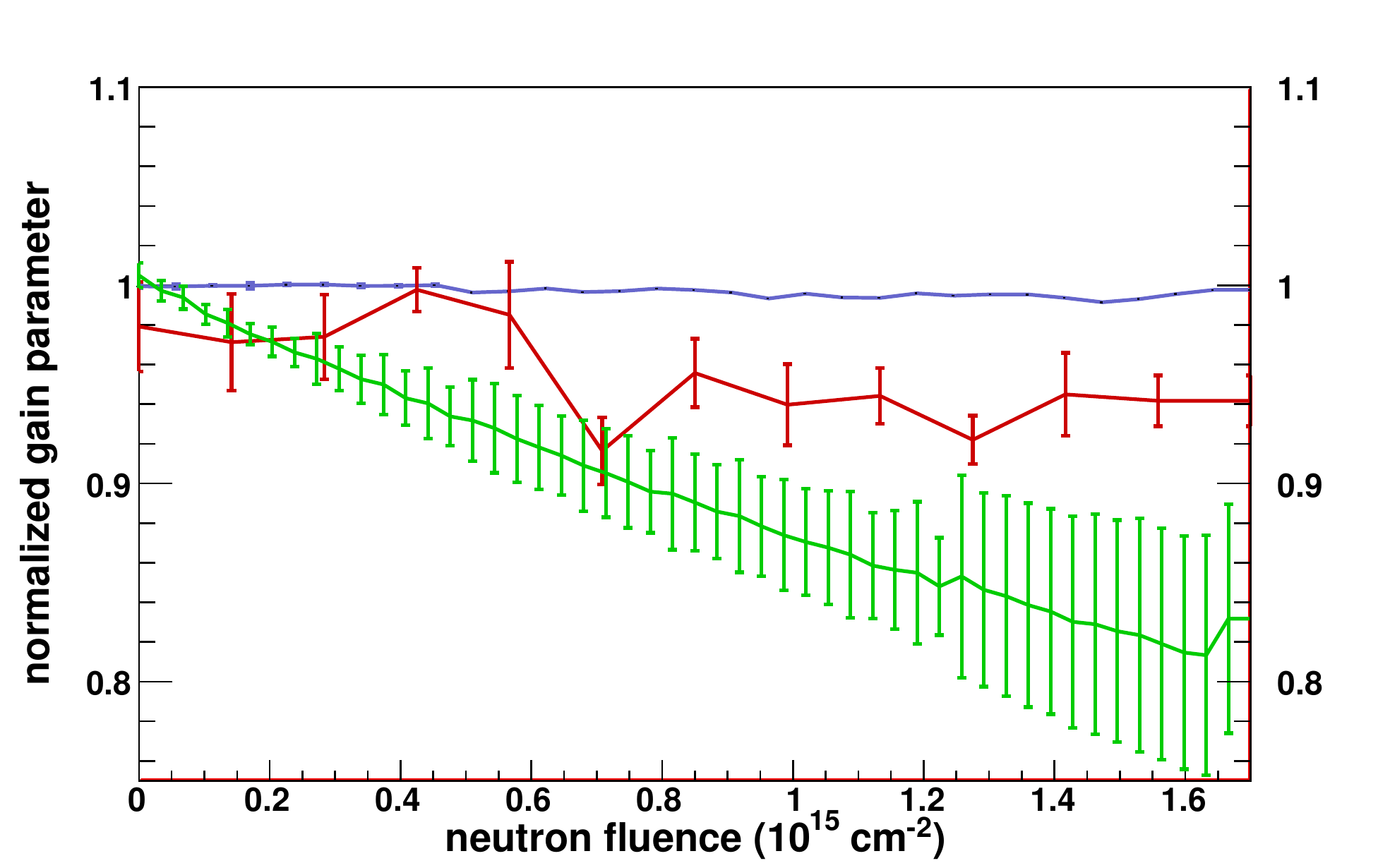}
\end{center}
\caption{Relative gain loss as a function of proton (top) and neutron (bottom) fluence for the various technologies described in the legend. The respective gain parameters are described in the text. The data points represent the average of all the devices of a particular technology that were irradiated up to a given particle fluence; the error bars are the sample RMS~\cite{5402221}.}
\end{figure}

The various transistor parameters were calculated from the measured
$S$-parameters using suitable small signal circuit models. The
transistor parameters were averaged over a certain frequency range to
obtain their mean and rms-values for every set of $S$-parameters. This
frequency range extended from $300\, \rm{kHz}$ to $100\, \rm{MHz}$,
unless a certain parameter was unstable at low or high 
frequencies, in which case appropriate cuts were applied. These averaged
transistor values were then used to characterize their behaviour as a
function of particle fluence.

Figure 3 summarizes the results for the proton (top) and the
neutron (bottom) tests in terms of the appropriate gain
parameters as a function of the corresponding particle fluence up to the required limit, normalized to the corresponding value before irradiation. The various gain parameters are the real 
part of the transconductance $g_m$ for the FET transistors, the differential
current gain $\beta$ for the Bipolar transistors, and in case of the HEC BB96 preamplifiers and systems, where our simple transistor models are not applicable, the transresistance $r_m = \frac{\Delta V_{\rm out}}{\Delta I_{\rm in}} = |S_{21} \cdot Z_{\rm in}|$, which relates the output voltage to the input current, where $Z_{\rm in}$ is the input impedance. Displayed are the sample averages of all the devices of a particular technology under test at each of the two test facilities, that were irradiated up to a given particle fluence; the error bars represent the sample RMS. The relative change of the 
gain parameter is quoted in Table 1, where we also used a linear extrapolation of the gain parameter up to the required fluence for those devices that did not reach that fluence.

\begin{table}\label{Result_table}
  \caption{Loss of gain of various transistor technologies under neutron and
    proton irradiation~\cite{5402221}.}
  \begin{center}
    \begin{tabular}{lcc}
      \br
      \multicolumn{1}{c}{Technology} & Neutron fluence & Proton fluence\\
      & $1.7 \cdot 10^{15} \, \mathrm{n} / \rm{cm}^{2}$ & $1.8 \cdot 10^{14} \,
      \mathrm{p} / \rm{cm}^{2}$\\
      \mr
      Si CMOS FET & 0 \% & -11 \%\\
      SiGe Bipolar & -5 \% & -7 \%\\
      GaAs FET & -20 \% & -17 \%\\
      \br
    \end{tabular}
  \end{center}
\end{table}

\section{Summary}

Our first observation is that proton irradiation seems to cause more damage to the electronics than neutron irradiation (at least for the Si CMOS and SiGe Bipolar transistors), despite the fact that an order of magnitude more neutrons than protons are expected to impact the HEC cold electronics. In the case of the GaAs FETs, the gain degradation is roughly the same for both types of irradiation, although a quantitative assessment suffers from the fact that two different gain parameters (transconductance vs. transresistance) are being compared.

It can be concluded that both the Si CMOS FETs and the SiGe Bipolar transistors considered as alternative technologies are more radiation hard than the currently used GaAs FETs. Among the alternative technologies, a preference would be given to the Si CMOS FETs, since the SiGe Bipolar transistors require a stabilization of their operation point, which would lead to a more complex circuitry to meet the same radiation hardness requirements. 

The GaAs technology currently used in the HEC cold electronics degrades significantly at the expected radiation levels, but it does not break down yet. The observed level of degradation ($\sim$20\%) could conceiveably be compensated for by some sort of calibration procedure. But before one can come to a decision whether it is safe or not to keep the current HEC cold electronics for HL-LHC conditions, the impact of this level of degradation on the overall HEC performance (e.g. in terms of jet energy resolution) needs to be better understood. In addition to that, a measurement of the performance degradation in the cold is needed. A decision has thus been reached to perform a more detailed study of the current HEC BB96 ASICs, including characterization of the linearity of their response with realistic pulses as they appear in the ATLAS detector, and a detailed simulation of the propagation of the 'degraded' signals through the summation, digitization, and reconstruction stages. This, and the measurement of the performance degradation in the cold are both ongoing efforts.

\section*{References}

\bibliographystyle{iopart-num}
\bibliography{calor2012}

\providecommand{\newblock}{}
\begin{thebibliography}{10}
\expandafter\ifx\csname url\endcsname\relax
  \def\url#1{{\tt #1}}\fi
\expandafter\ifx\csname urlprefix\endcsname\relax\def\urlprefix{URL }\fi
\providecommand{\eprint}[2][]{\url{#2}}

\bibitem{ATLAS}
 1994 {\em { ATLAS: technical proposal for a general-purpose pp experiment at
  the Large Hadron Collider at CERN }\/} LHC Tech. Proposal (Geneva: CERN)
  cern-lhcc-94-43, lhcc-p-2

\bibitem{JINST_3_S08003}
Aad G {\em et~al.\/} (ATLAS Collaboration) 2008 {\em J.Inst.\/} {\bf 3} S08003
  doi:10.1088/1748-0221/3/08/S08003

\bibitem{LArTDR}
{ATLAS Collaboration} (ATLAS) 1996 Atlas liquid argon calorimeter: Technical
  design report Tech. rep. CERN cern-lhcc-96-41, atlas tdr 2

\bibitem{MPP-2007-237}
Gingrich D~M {\em et~al.\/} (The ATLAS Hadronic End-Cap Calorimeter Group) 2007
  {\em J.Inst.\/} {\bf 2} P05005 atl-larg-pub-2007-009, atl-com-larg-2007-006,
  mpp-2007-237

\bibitem{MPP-2005-193}
Ban J, Brettel H, Cwienk W~D, Fent J, Kurchaninov L, Ladygin E, Oberlack H,
  Schacht P, Stenzel H and Strizenec P 2006 {\em Nucl.Instrum.Meth.A\/} {\bf
  556} 158--168 mpp-2005-193

\bibitem{Atlas-Upgrade}
Vankov P 2012 {\em arXiv:1201.5469\/} ATL-UPGRADE-PROC-2012-003

\bibitem{Schacht}
Schacht P 2010 {\em Proceedings of the 11th ICATPP Conference on Astroparticle,
  Particle, Space Physics, Detectors and Medical Physics Applications\/} (World
  Scientific) pp 392--403 ISBN 9789814307529 doi:10.1142/9789814307529\_0065

\bibitem{ATL-GEN-2005-001}
Baranov S, Bosman M, Dawson I, Hedberg V, Nisati A and Shupe M 2005 Estimation
  of radiation background, impact on detectors, activation and shielding
  optimization in atlas Tech. Rep. ATL-GEN-2005-001,
  ATL-COM-GEN-2005-001,CERN-ATL-GEN-2005-001 CERN Geneva (and unpublished
  updates)

\bibitem{SParameter}
Dicke R~H 1947 {\em J.Appl.Phys.\/} {\bf 18} 873 doi:10.1063/1.1697561

\bibitem{RadfilmCalib}
Marti\v{s}\'{i}kov\'{a} M and J\"{a}kel O 2010 {\em Phys.Med.Biol.\/} {\bf 55}
  N281--290 doi:10.1088/0031-9155/55/10/N03

\bibitem{NIEL}
Vasilescu A 1997 The niel scaling hypothesis applied to neutron spectra of
  irradiation facilities and in the atlas and cms sct Tech. Rep. ROSE/TN/97-2
  CERN

\bibitem{5402221}
Oberlack H, Dannheim D, Fischer A, Hambarzumjan A, Pospelov G, Reimann O,
  Rudert A and Schacht P 2009 {\em Nuclear Science Symposium Conference Record
  (NSS/MIC), 2009 IEEE\/} pp 758--762 ISSN 1095-7863

\end{thebibliography}

\end{document}